\documentclass[prc,showpacs,preprintnumbers,amsmath,amssymb]{revtex4}
\usepackage{graphicx}
\newcommand{\beq}{\begin{equation}}
\newcommand{\eeq}{\end{equation}}
\newcommand{\beqa}{\begin{eqnarray}}
\newcommand{\eeqa}{\end{eqnarray}}
\newcommand{\nn}{\nonumber}
\newcommand{\Sigs}{\Sigma_{\mathrm s} }
\newcommand{\Sigv}{\Sigma_{\mathrm v} }
\newcommand{\Sigo}{\Sigma_{\mathrm o} }

\newcommand{\kfj}{k_{\mathrm Fj} }
\newcommand{\bfgamma}{\mbox{\boldmath$\gamma$\unboldmath}}

\newcommand{\Sigsi}{\Sigma_{{\mathrm s},i} }
\newcommand{\Sigvi}{\Sigma_{{\mathrm v},i} }
\newcommand{\Sigoi}{\Sigma_{{\mathrm o},i} }
\begin{document}
\preprint{}
\title{Relativistic Description of Finite Nuclei Based on Realistic $NN$ Interactions.}
\author{E. N. E. van Dalen}
\email{eric.van-dalen@uni-tuebingen.de}
\author{H. M\"uther}
\affiliation{Institut f$\ddot{\textrm{u}}$r Theoretische Physik,
Universit$\ddot{\textrm{a}}$t T$\ddot{\textrm{u}}$bingen, Auf der
Morgenstelle 14, D-72076 T$\ddot{\textrm{u}}$bingen, Germany}
\begin{abstract}
A set of relativistic mean field models is constructed including the
Hartree and Hartree-Fock approximation accounting for the exchange of isoscalar
and isovector mesons as well as the pion. Density dependent coupling 
functions are determined to reproduce the components of the nucleon self-energy
at the Fermi surface, obtained within the Dirac-Brueckner-Hartree-Fock (DBHF) approach
using a realistic nucleon-nucleon interaction. It is investigated, to which
extend the various mean field models can reproduce the DBHF results for
the momentum dependence of the self-energies and the total energy of infinite 
matter. The mean field models are also used to evaluate the bulk properties of spherical
closed-shell nuclei. We find that the Hartree-Fock model allowing for the
exchange of $\sigma,\,\omega,\,\rho,\,\delta$ mesons and pions, yield the
best reproduction of the DBHF results in infinite matter and also provides a
good description of the properties of finite nuclei without any adjustment of
parameters.     

\end{abstract}
\maketitle
\section{Introduction}

One of the main challenges of theoretical nuclear physics is the attempt to
develop a microscopic theory, i.e. without any adjustment of parameters, for the
description of bulk properties of finite nuclei, which is based  on high
precision free space nucleon-nucleon ($NN$) interactions. This means that this
approach connects properties of baryons in the vacuum ($NN$ scattering data)
with nuclear systems at densities around the saturation density of nuclear
matter. Therefore, this approach should have  a high predictive power, when it
is used in extreme cases such as in highly isospin asymmetric nuclear systems.
The study of these exotic nuclear systems is a fast-growing field of physics.
The great scientific potential is demonstrated by the plans for large-scale
exotic-nuclear-beam facilities, such as the the future GSI facility in Germany,
and is driven by the expectation of observing  nuclear properties, which are
very different from those encountered so far, i.e. near the valley of stability.

However, most of these attempts have already failed in completing the first
milestone, which is the microscopic description of the saturation properties of
infinite nuclear matter in terms of such free space $NN$
interactions\cite{machl,baldo,polls,coester}. Therefore three-nucleon forces
have been introduced to fit the saturation point and/or properties of light
nuclei\cite{zuo,bogner}.  Relativistic many-body approaches, in particular the
Dirac-Brueckner-Hartree-Fock (DBHF) approach, however, have been successful in
describing the saturation properties of nuclear
matter~\cite{anastasio83,horow87,terhaar:1987,brockmann:1990,dejong:1998,
gross:1999,alonso:2003,vandalen:2004b,vandalen:2007,vandalen:2010a,vandalen:2010b}
without the necessity to introduce many-body forces. Therefore a significant
part of the 3-body terms,  which are required in non-relativistic
investigations, may represent the relativistic effects originating from the
so-called Z-graphs in the expansion of relativistic
propagators\cite{brown87,Bouyssy87}.

Although DBHF calculations have been quite successful in describing nuclear
matter, full Dirac Brueckner calculations are still too complex to allow an
application to finite nuclei at present, since the consistent treatment of
correlation  and relativistic effects for finite systems is a rather involved
problem. In fact, Van Giai et al.~\cite{giai2010} addressed this problem as one
of the main open problems in Nuclear Physics.   Different approximation schemes
have been developed, which treat either the relativistic effects or the
correlation effects in an approximative way. 

In the former approximation scheme, the Dirac effects  are treated via a kind of
local density approximation (LDA), whereas the correlations effects are taken
into account by solving the Bethe-Goldstone equation directly for the finite
nucleus under
consideration~\cite{muether:1988,muether:1990,fritz:1993,vandalen:2010b}.  This
means that the self-consistency requirements of a conventional BHF calculation
for a finite nucleus are satisfied, while the relativistic effects are taken
into account by evaluating the matrix elements of the potential in terms of
in-medium Dirac spinors.

In the latter approximation scheme, the Dirac effects are treated directly for
the finite nucleus, whereas the correlation effects are deduced from nuclear
matter via a local density approximation. This treatment of the correlation
effects can be accomplished by defining effective meson exchange interactions,
which are obtained from the Dirac-Brueckner self-energy components. Therefore,
the coupling constants of such an effective interaction are density dependent,
since  these self-energy components are density dependent~\cite{marcos:1989}.
In this way, a semi-phenomenological relativistic density functional can be
constructed. This approximation scheme to describe finite nuclei will be further
explored in this paper.

These semi-phenomenological relativistic density functionals can be divided into
two groups, semi-phenomenological density dependent relativistic Hartree (DDRH)
theories and semi-phenomenological density dependent relativistic Hartree-Fock
(DDRHF) theories.  However, between the semi-phenomenological DDRH theories and
the DBHF approach two essential differences exist concerning the structure of
the self-energy in nuclear matter~\cite{vandalen:2007,vandalen:2010b}. The first
difference is the absence of a spatial contribution of the vector self-energy
$\Sigma_V$ in the DDRH theory, which is present in the DBHF approach.  This
self-energy component originates from Fock exchange contributions which are not
present in the DDRH theory.  The second difference is that the DBHF self-energy
terms explicitly depend on the particle's momentum, a feature which is also
absent in the DDRH theory.  This momentum dependence reflects the non-locality
of the DBHF self-energy terms, which originates as well from the Fock exchange
terms as from non-localities in the underlying effective $NN$ interaction, the
$G$-matrix.  In the work of
Ref.~\cite{hofmann:2001,vandalen:2007,gogelein:2008}, renormalizations were
introduced to compensate for these  differences. However, a more fundamental
solution at the level of the structure of the self-energy is preferable and will
be investigated in this study.

An improvement of the simple mean field studies of the DDRH studies is the
density dependent relativistic Hartree-Fock  (DDRHF) theory with its inclusion
of the Fock terms, since the essential differences between  the DBHF approach
and the DDRH theory concerning the structure of the self-energy in nuclear
matter  discussed above may be cured by the inclusion of the Fock exchange
terms.  Due to the presence of Fock terms in the DDRHF theory, one obtains a
spatial contribution of the vector self-energy $\Sigma_V$ and momentum dependent
self-energy components as in the DBHF approach. However, first attempts of such 
DDRHF models based on microscopic approaches were not so successful as the DDRH
models~\cite{fritz:1994,shi:1995}. A possible reason is that only the isoscalar
coupling functions are density dependent and the other coupling functions, such
as of the $\pi$-meson and the $\rho$-meson, remain density independent in these
DDRHF theories~\cite{fritz:1994,shi:1995}. In addition, the $\delta$-meson is
absent in these DDRHF theories. However, in DDRH theories this coupling provides
a mechanism to account for the differences in the scalar self-energies, i.e. in
the corresponding effective Dirac masses, for neutrons and protons in isospin
asymmetric nuclear matter~\cite{schiller:2001}. Therefore, the DDRHF theories
can be further improved by including the isovector $\delta$-meson and extent the
density dependence to all coupling functions, which has not been done so far for
DDRHF theories based on microscopic calculations.

The aim of this study is not to provide an ``optimal'' DDRHF parameterization,
which could then be used in studies of finite nuclei. Instead we are
investigating various models, which are all fitted to reproduce the DBHF results
for the relativistic components of the nucleon self-energy at the Fermi surface.
We than compare these models in predicting other observables for nuclear matter
and finite nuclei. Therefore it is the aim to explore the limits and the
reliability of such DDRHF description to provide a reliable approximation scheme
for a Dirac-Brueckner-Hartree-Fock description of finite nuclei.

The plan of this paper is as follows. The relativistic DDRHF theory is discussed
in Sec.~\ref{sec:DDRHF}, which also introduces the various stages of the meson
exchange models considered and the resulting parameterization of the
density-dependent coupling constants. Results for the structure of infinite
matter and finite nuclei are presented and discussed in Sec.~\ref{sec:R}.
Finally, we end with a summary and the conclusion in Sec.~\ref{sec:S&C}.

\section{Density Dependent Relativistic Hartree-Fock.}
\label{sec:DDRHF}

A Lagrangian density of an interacting many-particle
system consisting of nucleons and mesons is the starting
point of a DDRHF theory. The Lagrangian density of the DDRHF theory presented here includes as well the isoscalar mesons $\sigma$ and $\omega$ as the isovector mesons $\delta$ and $\rho$. Furthermore, one has the pseudo-vector meson $\pi$. Therefore, the Lagrangian density consists of three parts:
the free baryon Lagrangian density $\mathcal{L}_B$,
the free meson Lagrangian density $\mathcal{L}_M$, and
the interaction Lagrangian density $\mathcal{L}_{\text{int}}$:
\begin{equation}\label{Lag_dens}
	\mathcal{L} = \mathcal{L}_B + \mathcal{L}_M + \mathcal{L}_{\text{int}},
\end{equation}
which takes on the explicit form
\begin{equation}
\mathcal{L}_B =   \bar{\Psi} ( \, i \gamma _\mu \partial^\mu - M ) \Psi,  
\end{equation}
\begin{equation}
  \mathcal{L}_M = {\textstyle \frac{1}{2}} \sum_{\iota= \sigma, \delta, \pi}
			\Big( \partial_\mu \Phi_\iota \partial^\mu \Phi_\iota - m_\iota^2 \Phi_\iota^2 \Big)   
  		 	- {\textstyle \frac{1}{2}} \sum_{\kappa = \omega, \rho}
			\Big( \textstyle{ \frac{1}{2}} F_{(\kappa) \mu \nu}\, F_{(\kappa)}^{\mu \nu}
				- m_\kappa^2 A_{(\kappa)\mu} A_{(\kappa)}^{\mu} \Big),      
\end{equation}
\begin{equation}
  \mathcal{L}_{\text{int}} =	- g_\sigma\bar{\Psi}  \Phi_\sigma \Psi
                - g_\delta \bar{\Psi}  \boldsymbol{\tau} \boldsymbol{\Phi}_\delta \Psi
-\frac{f_\pi }{m_\pi} \bar{\Psi}  \boldsymbol{\tau} \gamma_5 \gamma_\mu
			[\partial^\mu \boldsymbol{\Phi}_\pi]  \Psi
		 - g_\omega \bar{\Psi}  \gamma_\mu A_{(\omega)}^{ \mu } \Psi  
		- g_\rho \bar{\Psi}  \boldsymbol{\tau } \gamma_\mu  \boldsymbol{A}_{(\rho)}^{\mu } \Psi 
\label{eq:Lint}
\end{equation}
with the field strength tensor
$F_{(\kappa)\mu \nu} = \partial_{\mu} A_{(\kappa)\nu} - \partial_\nu A_{(\kappa)\mu}$
for the vector mesons. In the above Lagrangian density the nucleon field is denoted by $\Psi$
and the nucleon rest mass by $M$.
The scalar meson fields are $ \Phi_\sigma$ and $ \boldsymbol{\Phi}_\delta $ and
the vector meson fields are $ A_{(\omega)} $ and $ \boldsymbol{A}_{(\rho)} $. Furthermore, the pseudo-vector $\pi$ field is denoted by $\boldsymbol{\Phi}_\pi$. Moreover, the bold symbols denote vectors in the isospin space acting between the two species of nucleons.
The mesons have rest masses $m_\sigma$, $m_\omega$, $m_\delta$, $m_\rho$, and $m_\pi$
and couple to the nucleons with the strength of the coupling constants
$ g_\sigma$, $g_\delta$, $g_\omega$, $g_\rho$, and $f_\pi$.

\subsection{Nuclear Matter}

To obtain the field equations for the nucleons, we minimize the action
for variations of the fields $\bar{\Psi}$
included in the Lagrangian density of Eq.~(\ref{Lag_dens})
\begin{equation}
  \delta \int_{t_0}^{t_1} dt \int d^3x \,
      \mathcal{L}\big(\bar{\Psi}(x), \partial_\mu \bar{\Psi}(x), t \big) = 0.
\end{equation}
Finally the following field equation is obtained 
\begin{equation}
\frac{\partial}{\partial x^\mu}
\frac{\partial \mathcal{L} }{\partial(\partial_\mu \bar{\Psi})}
-  \frac{\partial \mathcal{L}}{\partial \bar{\Psi}}
	- \frac{\partial \mathcal{L}}{\partial \rho} \frac{\delta \rho}{\delta \bar{\Psi}}= 0.
\label{eq:ELE}
\end{equation}

Without density dependent meson-baryon vertices, Eq.~(\ref{eq:ELE}) 
reduces to the normal Euler-Lagrange field equation, since the third term vanishes. Evaluating Eq.~(\ref{eq:ELE}) for the  field $\bar{\Psi}$,
we obtain the Dirac equation. This Dirac equation for the nucleon field can be written as
\begin{equation}
\big(i\gamma_\mu \partial^\mu - M - ( \Sigma + \Sigma^{(r)} \gamma_0 )\big) \, \Psi = 0,
\label{eq:Dirac}
\end{equation}
where  $\Sigma$ is the nucleon self-energy and $\Sigma^{(r)}$ is the so-called rearrangement
contribution to the nucleon self-energy.

The self-energy $\Sigma$ is obtained as well for density dependent as density independent meson-baryon vertices 
and reads 
\begin{equation}\label{self_en_1}
\begin{split}
  \Sigma =  \Big( & g_\sigma \Phi_\sigma 
		+ g_\delta \boldsymbol{\tau} \boldsymbol{\Phi}_\delta 
		+ {\textstyle \frac{f_\pi }{m_\pi}} \boldsymbol{\tau} \gamma_5 \gamma_\mu 
			[\partial^\mu \boldsymbol{\Phi}_\pi]  
		+ g_\omega \gamma_\mu A_{(\omega)}^{ \mu }   \\
		& + g_\rho \boldsymbol{\tau } \gamma_\mu  \boldsymbol{A}_{(\rho)}^{\mu }  \Big).
\end{split}
\end{equation}
for the Lagrangian density in Eqs.~(\ref{Lag_dens})-(\ref{eq:Lint}). Furthermore, this nucleon self-energy can be split into different parts with well-defined
behavior under Lorentz transformations. Because of the requirement of translational and rotational invariance,
hermiticity, parity conservation, and time reversal invariance,
the most general form of the Lorentz structure of the self-energy is   
\beqa
\Sigma(k)= \Sigs (k) -\gamma_0 \, \Sigo (k) + 
\bfgamma  \cdot \textbf{k} \,\Sigv (k),
\label{subsec:SM;eq:self1}
\eeqa
where $\Sigs$, $\Sigo$, and $\Sigv$ components are Lorentz scalar
functions. Therefore, it is practical to define the following effective quantities
\begin{equation}
k^*=k(1+\Re e \Sigma_v(k)), 
\end{equation}
\begin{equation}
m^*(k) = M + \Re e \Sigma_s(k), 
\end{equation}
and
\begin{equation}
E^*(k)=E(k) + \Re e \Sigma_o(k).
\end{equation}
In the on-shell case, the effective energy can also be written as
\begin{equation}
E^*(k)^2={k^*}^2+m^*(k)^2.
\end{equation} 
On the level of the Hartree-Fock approximation, the contributions  to the self-energy of the Lagrangian density presented in Eqs.~(\ref{Lag_dens})-(\ref{eq:Lint}) are
\beqa
\Sigsi (k) & = & - \left(\frac{g_{\sigma}}{m_{\sigma}}\right)^2 (\rho_{s,n} + \rho_{s,p}) - \left(\frac{g_{\delta}}{m_{\delta}}\right)^2
\sum_{j=n,p} (\rho_{s,i}-\rho_{s,j})  
\nn \\ & &+ \frac{1}{(4 \pi)^2} 
\frac{1}{k} \sum_{j=n,p}
\int_0^{\kfj} q dq   \frac{m^*(q)}{E^*(q)}
\Bigg[ \delta_{ij} \left[ g_{\sigma}^2 \Theta_{\sigma}(k,q)  - 4 g_{\omega}^2 \Theta_{\omega}(k,q)  \right] 
\\ & &  + (2-\delta_{ij}) \left[-\left(\frac{f_\pi}{m_{\pi}}\right)^2 m_{\pi}^2 \Theta_{\pi}(k,q) - 4 g_{\rho}^2 \Theta_{\rho}(k,q) + g_{\delta}^2 \Theta_{\delta}(k,q)\right] \Bigg], \nn 
\eeqa
\beqa
\Sigoi (k) & = & -\left(\frac{g_{\omega}}{m_{\omega}}\right)^2 (\rho_n + \rho_p) - \left(\frac{g_{\rho}}{m_{\rho}}\right)^2 \sum_{j=n,p} (\rho_{i}-\rho_{j}) \nn \\
& & - \frac{1}{(4 \pi)^2} 
\frac{1}{k} \sum_{j=n,p}
\int_0^{\kfj} q dq \Bigg[ \delta_{ij} \left[ g_{\sigma}^2 \Theta_{\sigma}(k,q)  + 2 g_{\omega}^2 \Theta_{\omega}(k,q)  \right] \label{eq:Sigma0}
\\ & &   + (2-\delta_{ij}) \left[-\left(\frac{f_\pi}{m_{\pi}}\right)^2 m_{\pi}^2 \Theta_{\pi}(k,q) + 2 g_{\rho}^2 \Theta_{\rho}(k,q) + g_{\delta}^2 \Theta_{\delta}(k,q)\right] \Bigg], \nn 
\eeqa
and
\beqa
\Sigvi (k) & = & - \frac{1}{(4 \pi k)^2}   \sum_{j=n,p}
\int_0^{\kfj}  dq \frac{q^*}{E^*(q)}
\Bigg[\delta_{ij} \left[ 2 g_{\sigma}^2 \Gamma_{\sigma}(k,q) + 4  g_{\omega}^2 \Gamma_{\omega}(k,q) \right] \nn \\  
 & & \quad + (2-\delta_{ij}) \Bigg[-2 \left(\frac{f_{\pi}}{m_{\pi}}\right)^2 \left( (k^2+q^2) \Gamma_{\pi}(k,q) - k q^2 \Theta_\pi(k,q) \right)
+ 4 g_{\rho}^2 \Gamma_{\rho}(k,q) \nn\\
&&\quad + 2 g_{\delta}^2 \Gamma_{\delta}(k,q) \Bigg] \Bigg].
\eeqa
The first two terms in $\Sigsi$ and $\Sigoi$ correspond to the Hartree contribution with
\begin{eqnarray}
\rho_{s,i}& = &\frac{1}{\pi^2} \int_0^{k_{Fi}} q^2 dq \frac{m^*(q)}{E^*(q)}, \\
\rho_{i} & = & \frac{k_{Fi}^3}{3 \pi^2}
\end{eqnarray} 
The remaining expressions are due to the Fock contributions, where the abbreviations
\begin{equation}
A(k,q)=q^2+k^2-\left(E(k)-E(q)\right)^2,
\end{equation}
\begin{equation}
\Theta_i(k,q)=\ln\left(\frac{A(k,q)+m_i^2 + 2 k q}{A(k,q)+m_i^2 - 2 k q}\right),
\end{equation}
and
\begin{equation}
\Gamma_i(k,q)=\frac{\left((A(k,q)+m_i^2) \Theta_i(k,q)\right)}{4 k}-q
\end{equation}
are used.

The $\Sigma^{(r)}$ term in Eq.~(\ref{eq:Dirac}) will only be present, if meson-baryon vertices are density dependent, and is generated by the third term in Eq.~(\ref{eq:ELE}). This rearrangement contribution $\Sigma^{(r)}$ reads
\begin{equation}\label{self_en_rearr_1}
\begin{split}
  \Sigma^{(r)} = \Big(
	& \frac{\partial g_\sigma}{\partial \rho} \bar{\Psi} \Phi_\sigma \Psi
	 + \frac{\partial g_\delta}{\partial \rho} \bar{\Psi} \boldsymbol{\tau} \boldsymbol{\Phi}_\delta \Psi
+  \frac{1}{m_\pi}\frac{f_\pi}{\partial \rho} \bar{\Psi} \boldsymbol{\tau} \gamma_5 \gamma_\mu 
			[\partial^\mu \boldsymbol{\Phi}_\pi] \Psi  \\ &
	 + \frac{\partial g_\omega}{\partial \rho} \bar{\Psi} \gamma_\mu A_{(\omega)}^{ \mu } \Psi 
	  + \frac{\partial g_\rho}{\partial \rho}
	      \bar{\Psi} \boldsymbol{\tau } \gamma_\mu  \boldsymbol{A}_{(\rho)}^{\mu } \Psi
            \Big).
\end{split}
\end{equation}
These rearrangement contributions are essential
to provide a symmetry conserving approach, which implies that energy-momentum 
conservation and thermodynamic consistency like the Hugenholtz - van Hove
theorem are satisfied\cite{fuchs:1995}.  However, these rearrangement terms do not 
contribute to the energy per nucleon,
\begin{equation}
E/A = T + V - M,
\label{eq:EoverA}
\end{equation}
where the kinetic energy per nucleon is
\begin{equation}
T=\left[M \rho_s + \frac{1}{\pi^2} \sum_{i=n,p} \int_0^{k_{Fi}} \frac{q^*}{E^*(q)} q^3 dq \right] \frac{1}{\rho_B} 
\label{eq:Ekin}
\end{equation}   
and the potential energy per nucleon is 
\begin{equation}
V= \frac{1}{2 \pi^2 \rho_B} \sum_{i=n,p}  \int_0^{k_{Fi}} \left(\frac{m^*(q)}{E^*(q)} \Sigsi(q) - \Sigoi (q) + \frac{q^*}{E^*(q)} q \Sigvi(q)
\right) q^2 dq, 
\label{eq:potential}
\end{equation}
with $\rho_s=\rho_{s,n}+\rho_{s,p}$ and $\rho_B=\rho_n+\rho_p$. 

Furthermore, the rearrangement terms can be added to the time-like vector self-energy in Eq.~(\ref{eq:Sigma0}). Therefore, it should be noted that these rearrangement terms are sometimes included in the definition of  the time-like vector self-energy $\Sigma_0 (k)$. 
\subsection{Parameterization}
We have constructed three different DDRHF models: a $\sigma\omega$ model, a
$\sigma\omega\pi$ model, and a $\sigma\omega\pi\rho\delta$ model. The effective coupling
constants for the mesons are determined  by requesting that the HF expression for the scalar
self-energy  $\Sigma_s (k)$ and the time-like vector self-energy $\Sigma_0 (k)$, i.e. without
the rearrangement terms $\Sigma^{(r)}$, calculated  at the Fermi surface reproduce the
corresponding results of a DBHF calculation using Bonn A~\cite{vandalen:2007}.  The reason that
the rearrangement terms are not included in the fit is that the DBHF approach has no
rearrangement contributions. The DBHF data at densities of $\rho_B=0.100 , 0.197, 0.313$, and
$0.467$ fm$^{-3}$ are used to obtain the parameters of the various models. The density
dependent couplings of the $\sigma\omega$ and  the $\sigma\omega\pi$ model  are obtained
from the DBHF results in isospin symmetric nuclear matter ($Y_p=0.5$), whereas couplings of the
$\sigma\omega\pi\rho\delta$ model are determined from results in isospin asymmetric nuclear
matter with a proton fraction of  $Y_p=0.4$. For completeness two different Hartree (DDRH) models
have been constructed in a similar way.

\begin{table}
\begin{center}
\begin{tabular}{|c|c|cccc|}
\hline
\ \ \ meson i & \ $m$ [MeV]\ \ \ &\ \ \ $a_i$\ \ \ &\ \ \ $b_i$ \ \ \ 
& \ \ \ $c_i$  \ & \ \ \ $d_i$ \ \ \  	\\
\hline
 $\sigma$ & 550 & $9.28996$     & $-0.8403$    & $0.25791$   &  $-0.015436$   \\
 $\omega$ & 782.6 & $10.23446$    & $0.23622$    & $-0.1503$   &  $0.0297512$   \\
\hline
 $\sigma$ & 550 & $8.86711$     & $-1.1508$    & $0.39224$   &  $-0.034995$   \\
 $\omega$ & 782.6 & $8.37380$    & $0.71233$    & $-0.2986$   &  $0.0469882$   \\
\hline

\end{tabular}
\end{center}
\caption{\label{table:DDRHF_parameters_sw} Parameter set for the $\sigma\omega$ model 
from the DBHF approach in Ref.~\cite{vandalen:2007}. The upper part of the table contains
the parameters to be used in the Hartree approach, whereas the lower part refers to the
DDRHF model.}
\end{table}

\begin{table}
\begin{center}
\begin{tabular}{|c|c|cccc|}
\hline
\ \ \ meson i & \ $m$ [MeV]\ \ \ &\ \ \ $a_i$\ \ \ &\ \ \ $b_i$ \ \ \ 
& \ \ \ $c_i$  \ & \ \ \ $d_i$ \ \ \  	\\
\hline
 $\sigma$ & 550 & $8.65683$     & $-1.0265$    & $0.35362$   &  $-0.0408585$   \\
 $\omega$ & 782.6 & $8.62502$    & $0.53724$    & $-0.2312$   &   $0.0386465$   \\
 $\pi$ & 139 & $1.00265$   &   -  &  -   &  - \\
\hline
\end{tabular}
\end{center}
\caption{\label{table:DDRHF_parameters_swp}Parameter set for the $\sigma\omega\pi$ DDRHF model  from the DBHF approach in Ref.~\cite{vandalen:2007}.}
\end{table}

\begin{table}
\begin{center}
\begin{tabular}{|c|c|cccc|}
\hline
\ \ \ meson i & \ $m$ [MeV]\ \ \ &\ \ \ $a_i$\ \ \ &\ \ \ $b_i$ \ \ \ 
& \ \ \ $c_i$  \ & \ \ \ $d_i$ \ \ \  	\\
\hline
 $\sigma$ & 550 & $9.26408$     & $-0.79477$    & $0.24135$   &  $-0.013504$   \\
 $\omega$ & 782.6 & $10.18505$    & $0.31476$    & $-0.19474$   & $0.037339$    \\
 $\delta$ & 983 & $9.58644$   & $-4.34658$    & $1.56009$    & $-0.208196$ \\
 $\rho$   & 769   & $9.51803$    & $-3.36524$    & $1.36753$   & $0.158856$     \\
\hline
 $\sigma$ & 550 & $4.99231$     & $3.27705$    & $-1.4929$   &  $0.230977$   \\
 $\omega$ & 782.6 & $0.67868$    & $10.3068$    & $-4.6471$   & $0.701354$    \\
   $\pi$   & 139 &  1.00265   &   -  &   -   &   -   \\
 $\delta$ & 983 & $12.3664$   & $-8.9781$    & $3.25210$    & $-0.426997$ \\
 $\rho$   & 769   & $15.2464$    & $-11.1504$    & $4.27083$   & $-0.543179$     \\
\hline
\end{tabular}
\end{center}
\caption{\label{table:DDRHF_parameters_swprd} Parameter set for the 
$\sigma\omega\pi\rho\delta$ model from the DBHF approach in 
Ref.~\cite{vandalen:2007}. The upper part of the table contains
the parameters to be used in the Hartree approach, whereas the lower part refers to the
DDRHF model. Note that the $\pi$ does not contribute in the Hartree approximation.}
\end{table}

In order to make these parameterizations easily accessible, we have parameterized 
the density dependence of the coupling constants by
\begin{equation}
  g_i (\rho _B)
    = a_i + b_i x + c_i  x^2 + d_i x^3 ,
\label{eq:EvD_coupl_func}
\end{equation}
with $x = \rho _B/\rho_0$ and $\rho_0$ = 0.16 fm$^{-3}$. The parameters of the coupling functions are fitted, except the one of the $\pi$ coupling function. The $\pi$ coupling constant is fixed to the free value. In addition, the masses of the mesons
are chosen to be identical to those of the Bonn A potential. All parameters  are summarized in table \ref{table:DDRHF_parameters_sw} for the 
$\sigma\omega$ model, 
in table \ref{table:DDRHF_parameters_swp} for  the $\sigma\omega\pi$ model, and in table \ref{table:DDRHF_parameters_swprd} for  the 
$\sigma\omega\pi\rho\delta$ model.
\begin{figure}[!t]
\begin{center}
\includegraphics[width=0.48\textwidth] {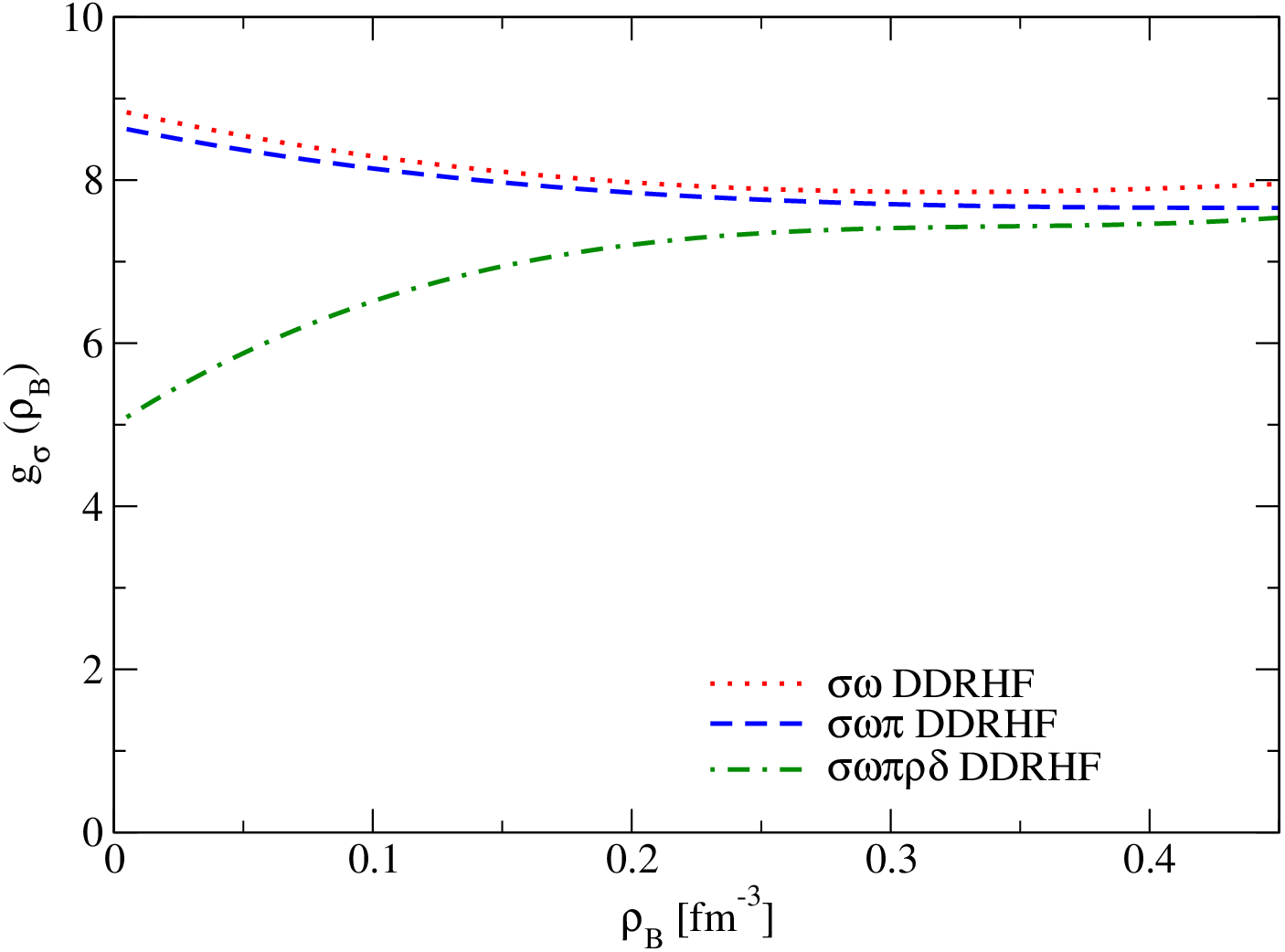} \quad
\includegraphics[width=0.48\textwidth] {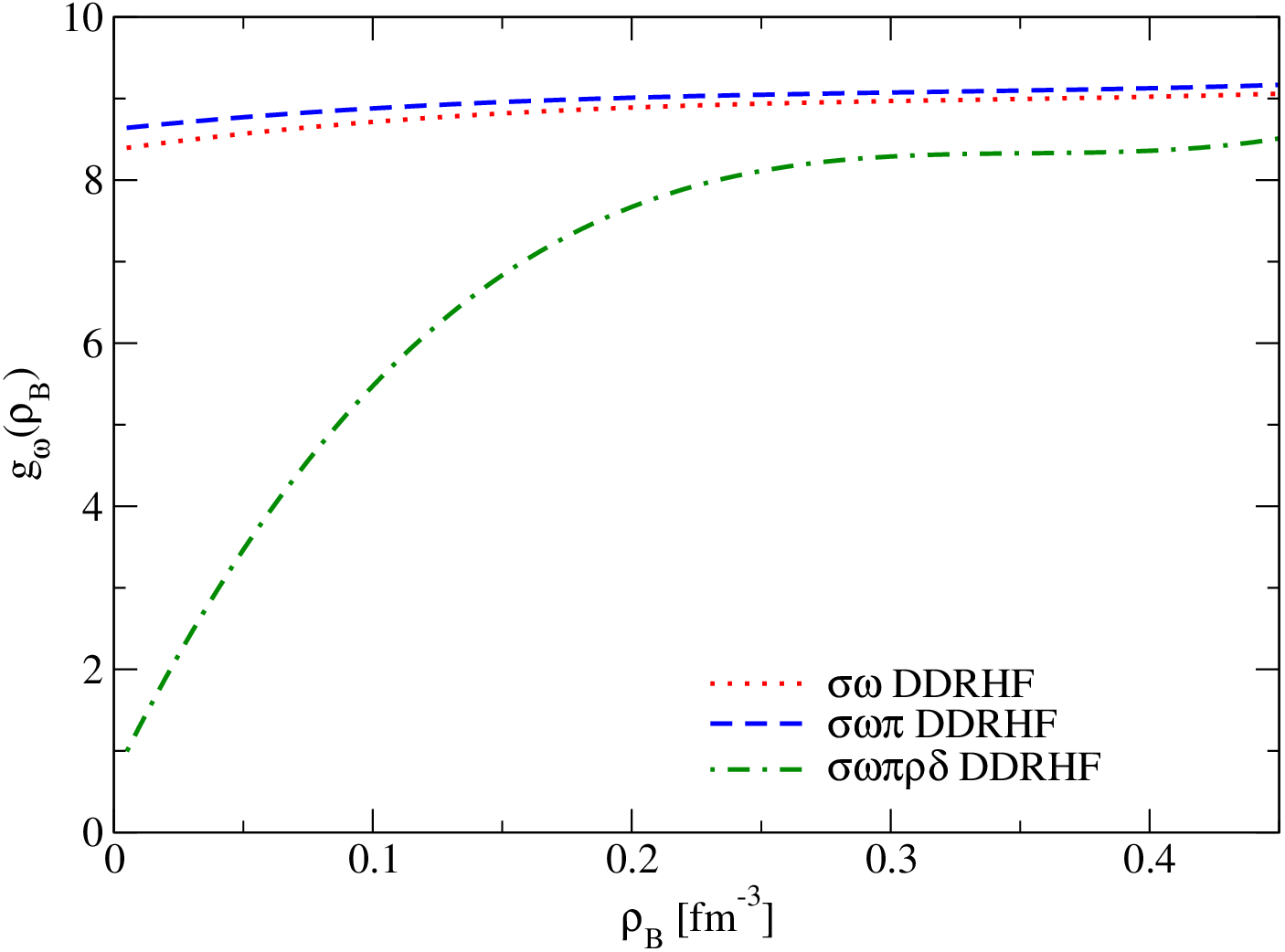}
\caption{(Color online) Density dependence of the isoscalar coupling functions of the $\sigma\omega$ (dotted), $\sigma\omega\pi$(dashed), and the $\sigma\omega\pi\rho\delta$(dashed-dotted) DDRHF model. Left: the $\sigma$ coupling function. Right: the $\omega$ coupling function.}
\label{fig:isoscalar}
\end{center}
\end{figure}

The density dependence of the isoscalar coupling functions is displayed in
Fig.~\ref{fig:isoscalar}. In the $\sigma\omega$ model, the $\sigma$ coupling
function is decreasing with increasing density, whereas the $\omega$ coupling
function is slightly increasing with increasing density. These features can be interpreted in the following way: 

A significant part of the medium-range attraction contained in the Brueckner
$G$-matrix is due to the  iterated $\pi$-exchange term. Pauli blocking and
dispersive corrections of the nucleon-nucleon propagator in the nuclear medium
yield a quenching of these iterated $\pi-$exchange terms with increasing density
(see e.g. \cite{machl} and \cite{polls}). In the $\sigma\omega$ model this
medium range attraction is described in terms of the $\sigma$-exchange.
Therefore, the quenching of the iterated  $\pi$-exchange terms leads to a
reduction of the coupling constant for the $\sigma$ with increasing density.

The same Pauli blocking effects and dispersive corrections in the NN propagator also reduce the
correlation effects in the relative wave functions of two nucleons at short distances.
Therefore, these short-range correlations are less efficient at higher densities to minimize the 
repulsive short-range components of the bare $NN$ interaction. This leads to a slight increase
of the effective coupling constant for the $\omega$ with increasing density.

Next, we will compare the isoscalar coupling functions  of the $\sigma\omega$ model with that
of the $\sigma\omega\pi$ model. One finds that the inclusion of the $\pi$ exchange has a
small effect on the isoscalar coupling functions, i.e. the $\sigma$ coupling function is
slightly decreased, whereas the  $\omega$ coupling function is slightly enhanced compared to
the $\sigma\omega$ model. 

\begin{figure}[!t]
\begin{center}
\includegraphics[width=0.48\textwidth] {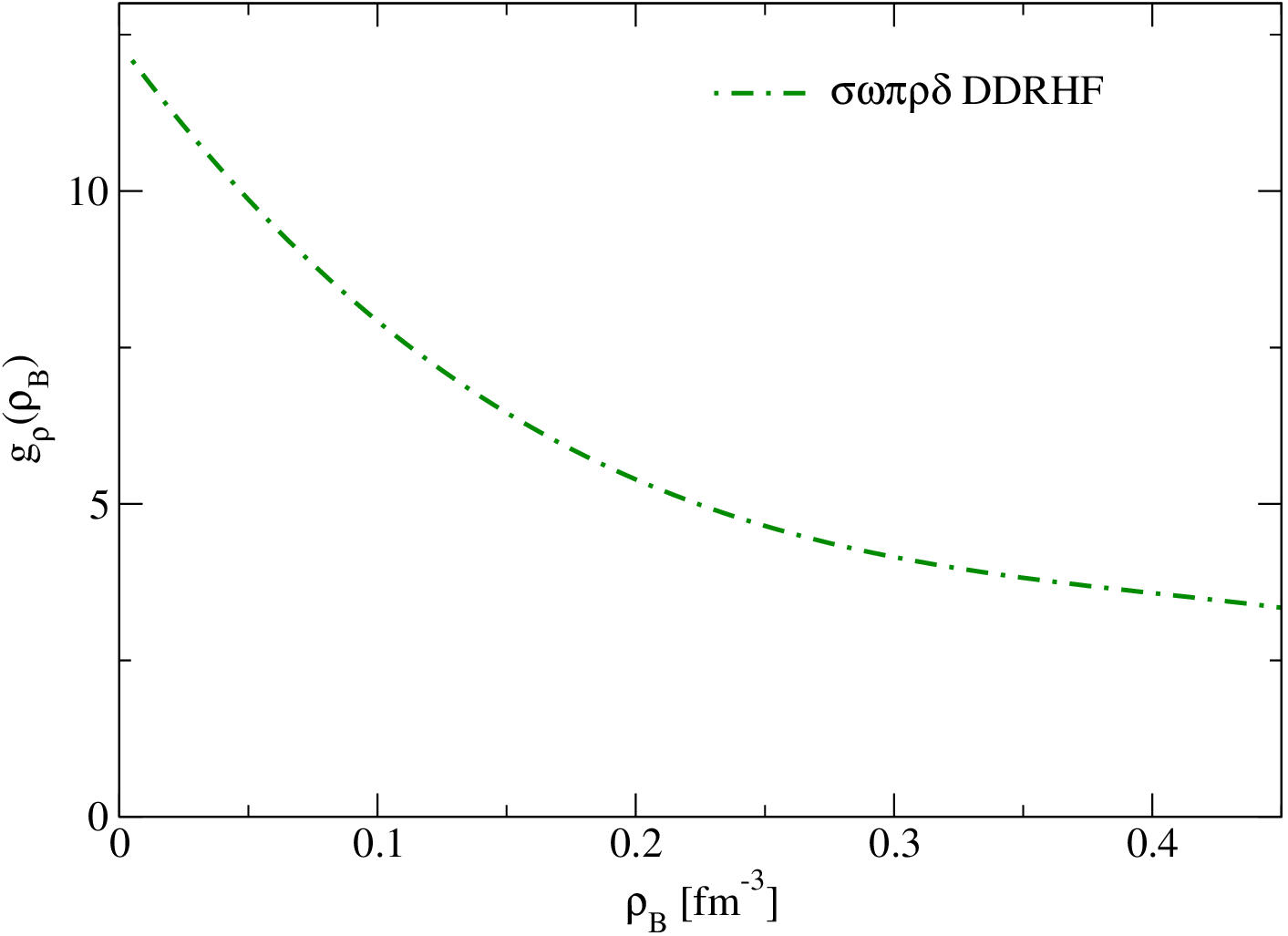} \quad
\includegraphics[width=0.48\textwidth] {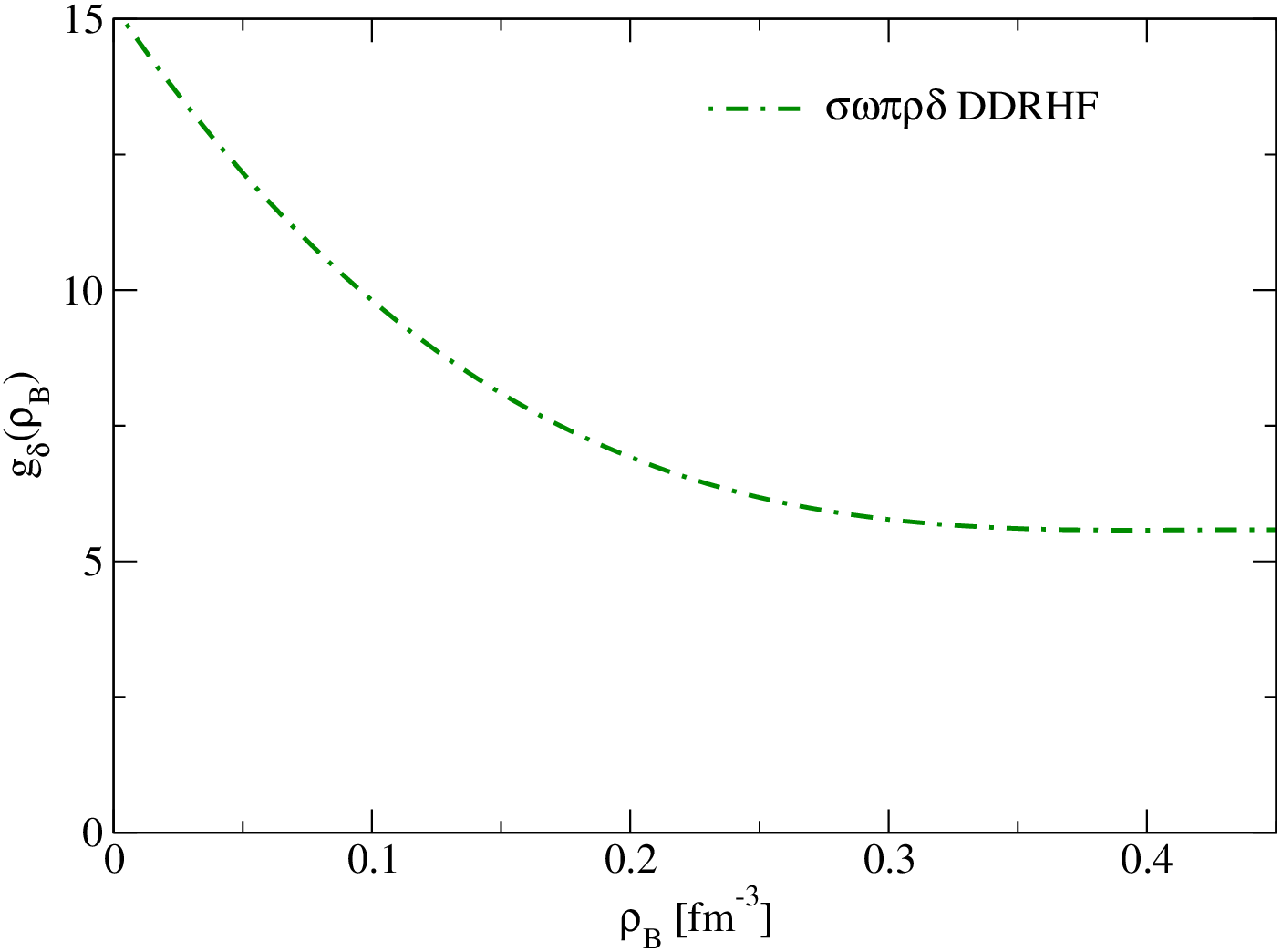}
\caption{(Color online) Density dependence of the isovector coupling functions of the $\sigma\omega\pi\rho\delta$ DDRHF model. Left: the $\rho$ coupling function. Right: the $\delta$ coupling function.}
\label{fig:isovector}
\end{center}
\end{figure}

However, in the  $\sigma\omega\pi\rho\delta$ model the density dependence of
the isoscalar coupling functions dramatically change compared to the 
$\sigma\omega$ model and the $\sigma\omega\pi$ model. The effective coupling
constants for the isovector mesons are rather large in particular at small
densities as displayed in Fig.~\ref{fig:isovector}. This reflects the
significant differences between the isospin $T=1$ and $T=0$ interaction, which
is required to describe nucleon self-energies in isospin asymmetric nuclear
matter. The resulting interplay between the exchange terms for isoscalar and
isovector mesons spoils the simple picture to explain the density dependence of
the coupling constants for the scalar mesons discussed above. 
\section{Results and Discussion}
\label{sec:R}
\subsection{Nuclear Matter}
In this section we will present the nuclear matter results of the previously
constructed DDRHF models and compare them to the original DBHF approach of 
Ref.~\cite{vandalen:2007}, which the various parameterizations attempt to
reproduce.
First, the momentum dependence of the DDRHF self-energy components will be
considered. Note that the parameterizations are fitted to the self-energy
components at the Fermi momentum $k_F$. The momentum dependence of the DDRHF
self-energies exclusively originates from the various Fock exchange terms, whereas the
momentum dependence of the original DBHF self-energies is due to Fock exchange
terms but also due to the non-locality and energy-dependence of the underlying
$G$-matrix.

\begin{figure}[!t]
\begin{center}
\includegraphics[width=0.48\textwidth] {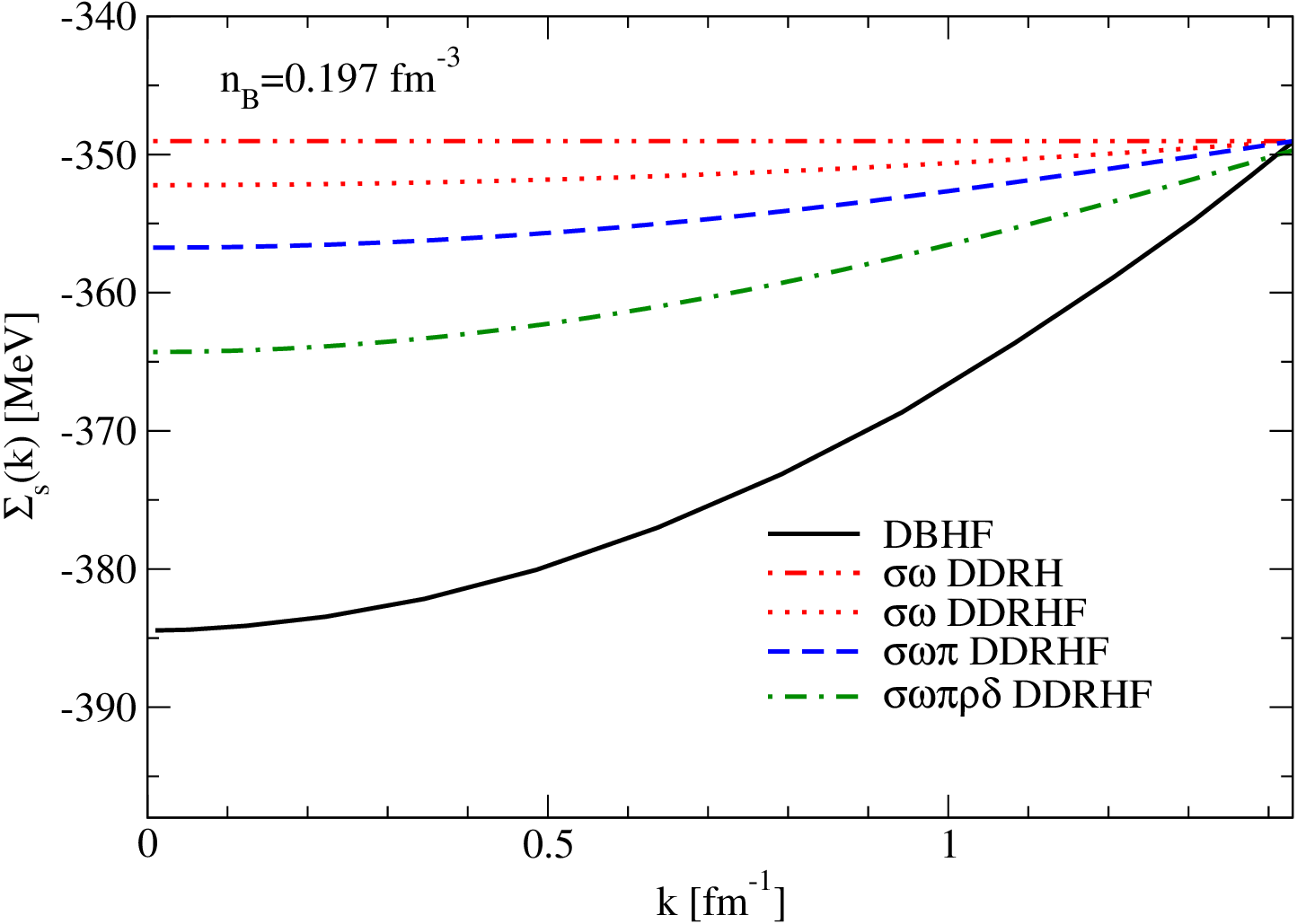} \quad
\includegraphics[width=0.48\textwidth] {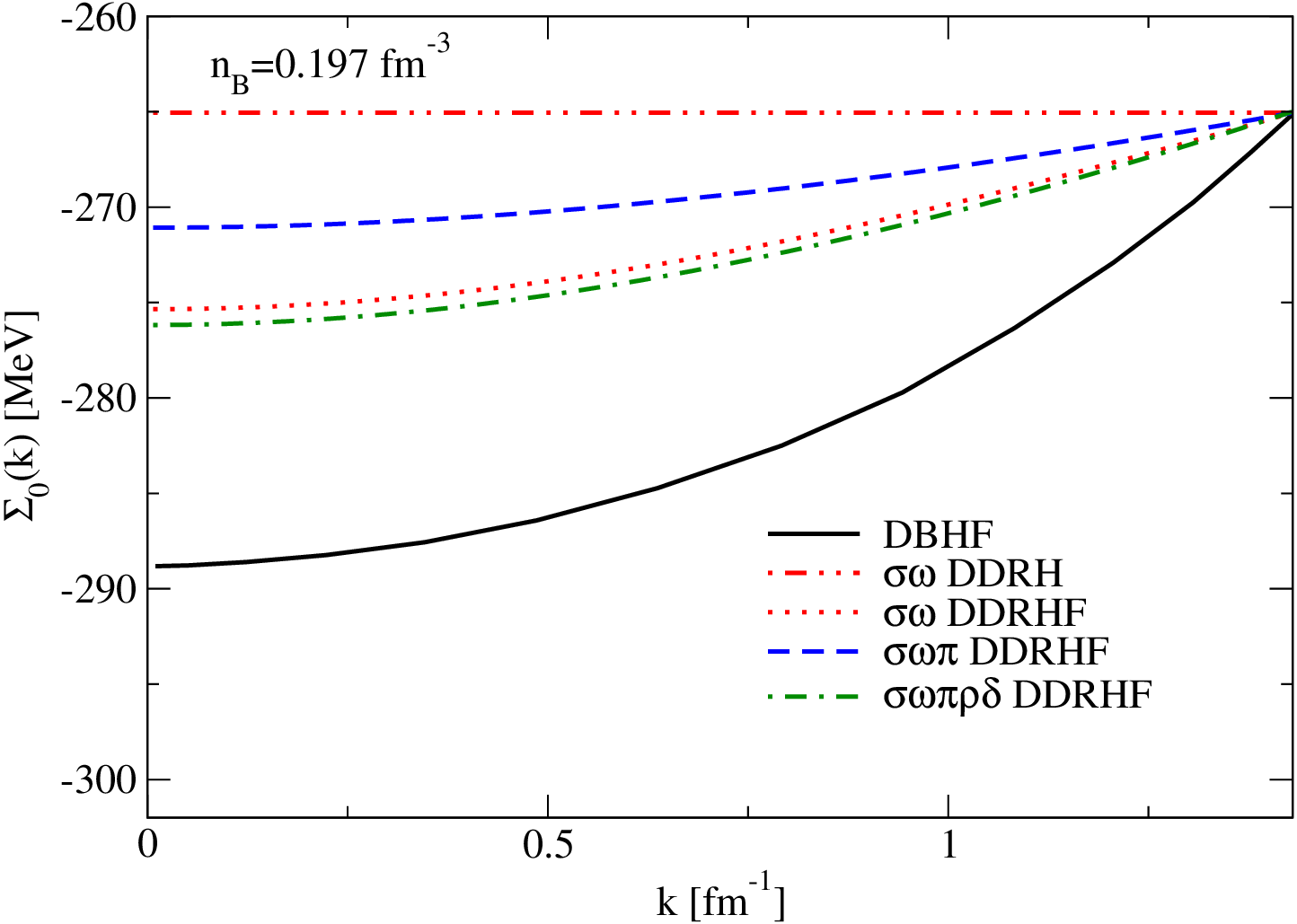}
\caption{(Color online) Momentum dependence of the scalar self-energy $\Sigma_s$ and the time-like vector self-energy $\Sigma_0$  for the $\sigma\omega$ (dotted), $\sigma\omega\pi$(dashed), and the $\sigma\omega\pi\rho\delta$(dashed-dotted) DDRHF model is plotted at a density of $\rho_B=0.197$ fm$^{-3}$. In addition, the Hartree model with $\sigma$ and $\omega$ mesons (dashed-dotted dotted) is given. Furthermore, the corresponding self-energy components of the DBHF approach in Ref.~\cite{vandalen:2007} (solid line) are presented in this figure. Left: the scalar self-energy $\Sigma_s$. Right: the time-like vector self-energy $\Sigma_0$.}
\label{fig:momselfenergy}
\end{center}
\end{figure}
This momentum dependence of the self-energy components in isospin symmetric nuclear
matter of a density $\rho_B=0.197$ fm$^{-3}$ for the various DDRHF models is
displayed in Fig.~\ref{fig:momselfenergy}. The Hartree approximation ignores all
Fock exchange terms and therefore provides a scalar self-energy which is independent
on the nucleon momentum $k$. Also for  the $\sigma\omega$ DDRHF model, the scalar
self-energy $\Sigma_s$ turns out to be almost momentum independent. The inclusion of the $\pi$
improves the momentum dependence of the scalar self-energy $\Sigma_s$. 

However, for
the  time-like vector self-energy $\Sigma_0$ the opposite behavior can be observed,
i.e. the inclusion of the $\pi$  deteriorates the results. Therefore, a
$\sigma\omega\pi$ DDRHF model will in principle not be able to reproduce the
momentum dependence of the DBHF self-energy components. Although the momentum
dependence in the  $\sigma\omega\pi\rho\delta$ model is in better agreement with
the DBHF approach than  in the other two models, its momentum dependence is only
about one half of the momentum dependence observed in the DBHF approach. 

As the models are fitted to reproduce the values of the the self-energy components 
at the Fermi momentum $k=k_F=1.43$ fm$^{-3}$ all models converge 
at this point. The  $\sigma\omega\pi\rho\delta$ DDRHF model is
fitted in isospin asymmetric nuclear matter, therefore a small deviation for this DDRHF model
can be observed in isospin symmetric nuclear matter. 

\begin{figure}[!h]
\begin{center}
\includegraphics[width=0.6\textwidth] {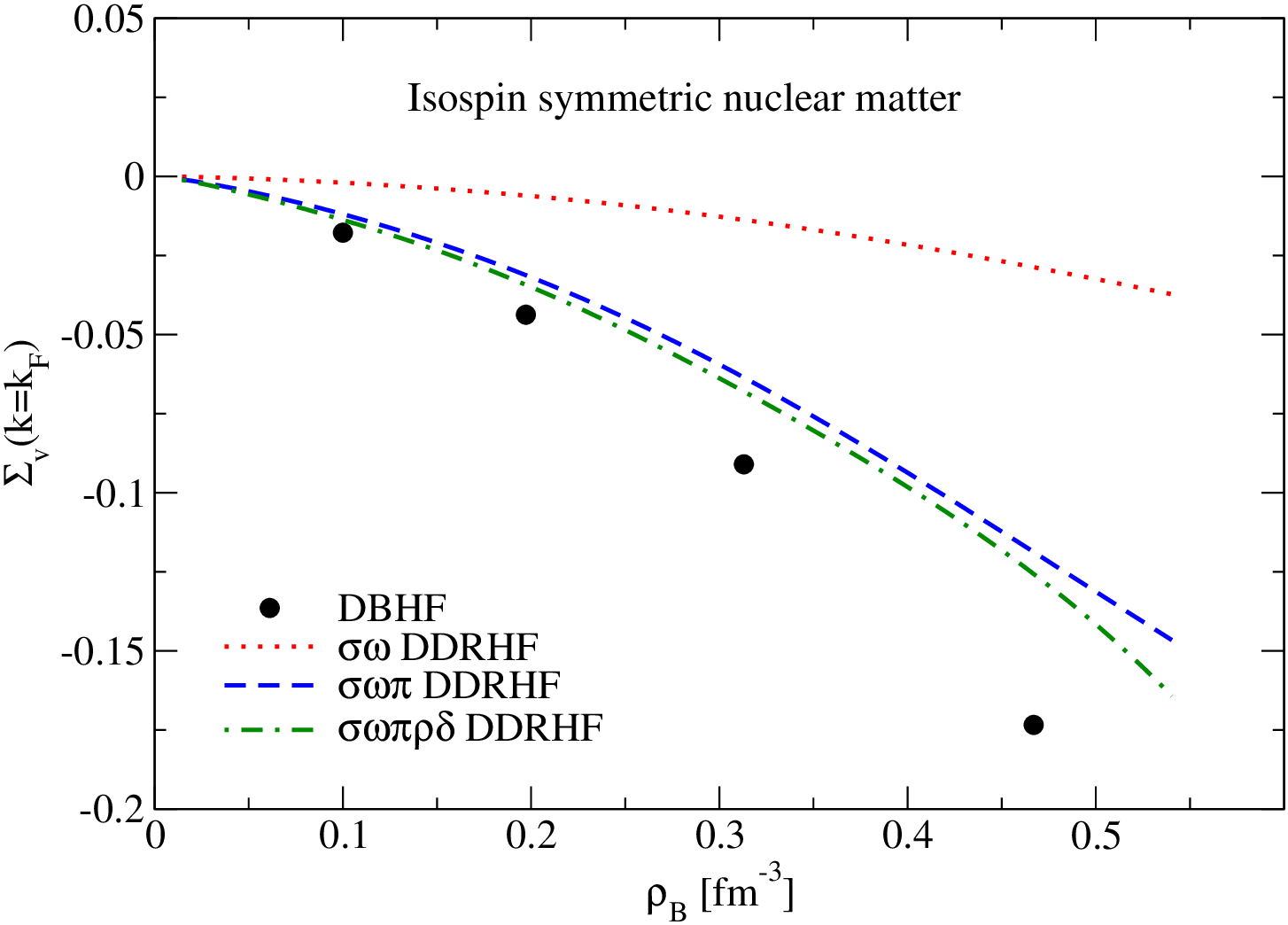}
\caption{(Color online) Density dependence of the spatial vector self-energy $\Sigma_v$  for the $\sigma\omega$ (dotted), $\sigma\omega\pi$(dashed), and the $\sigma\omega\pi\rho\delta$(dashed-dotted) DDRHF model is plotted in isospin symmetric nuclear matter. In addition, the corresponding self-energy component of the DBHF approach in Ref.~\cite{vandalen:2007} is presented in this figure.}
\label{fig:spavector}
\end{center}
\end{figure}

The other essential difference between the relativistic Hartree model DDRH and the DBHF approach is the
absence of a spatial contribution of the vector self-energy $\Sigma_V$ in the DDRH
theory. Such a contribution is present in the DDRHF theory due to the Fock exchange
terms.   In our procedure to determine the various DDRHF models this
self-energy component is not fitted. Results of various parameterizations are
displayed in Fig.~\ref{fig:spavector}. One observes that the models with $\pi$ 
exchange yield a stronger spatial component and show a much  better agreement with 
the DBHF result than the $\sigma\omega$ model. Therefore, it can be concluded that the inclusion of the
$\pi$  exchange in DDRHF models is essential in reproducing the spatial vector
self-energy.

\begin{figure}[!h]
\begin{center}
\includegraphics[width=0.48\textwidth] {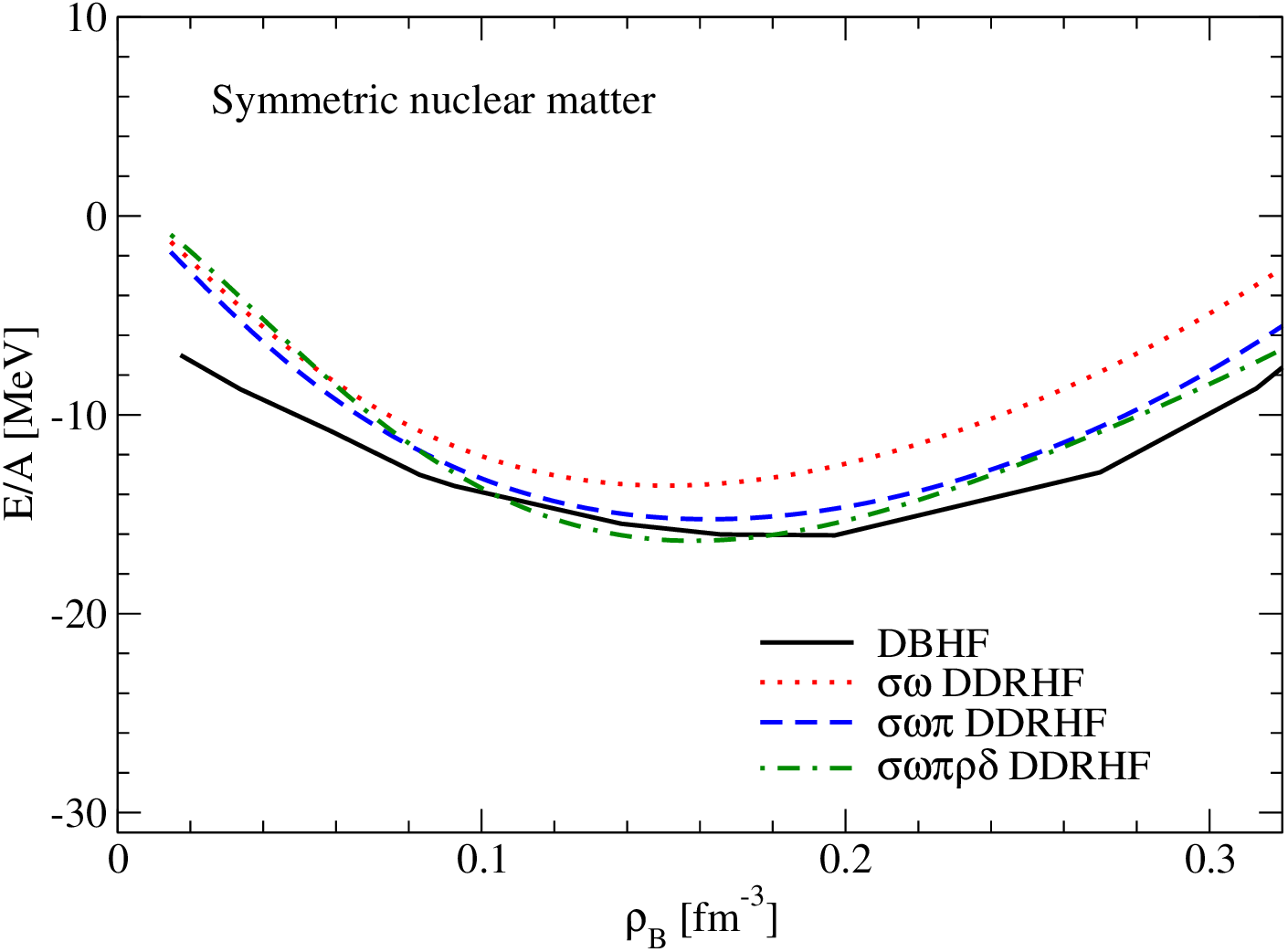} \quad
\includegraphics[width=0.48\textwidth] {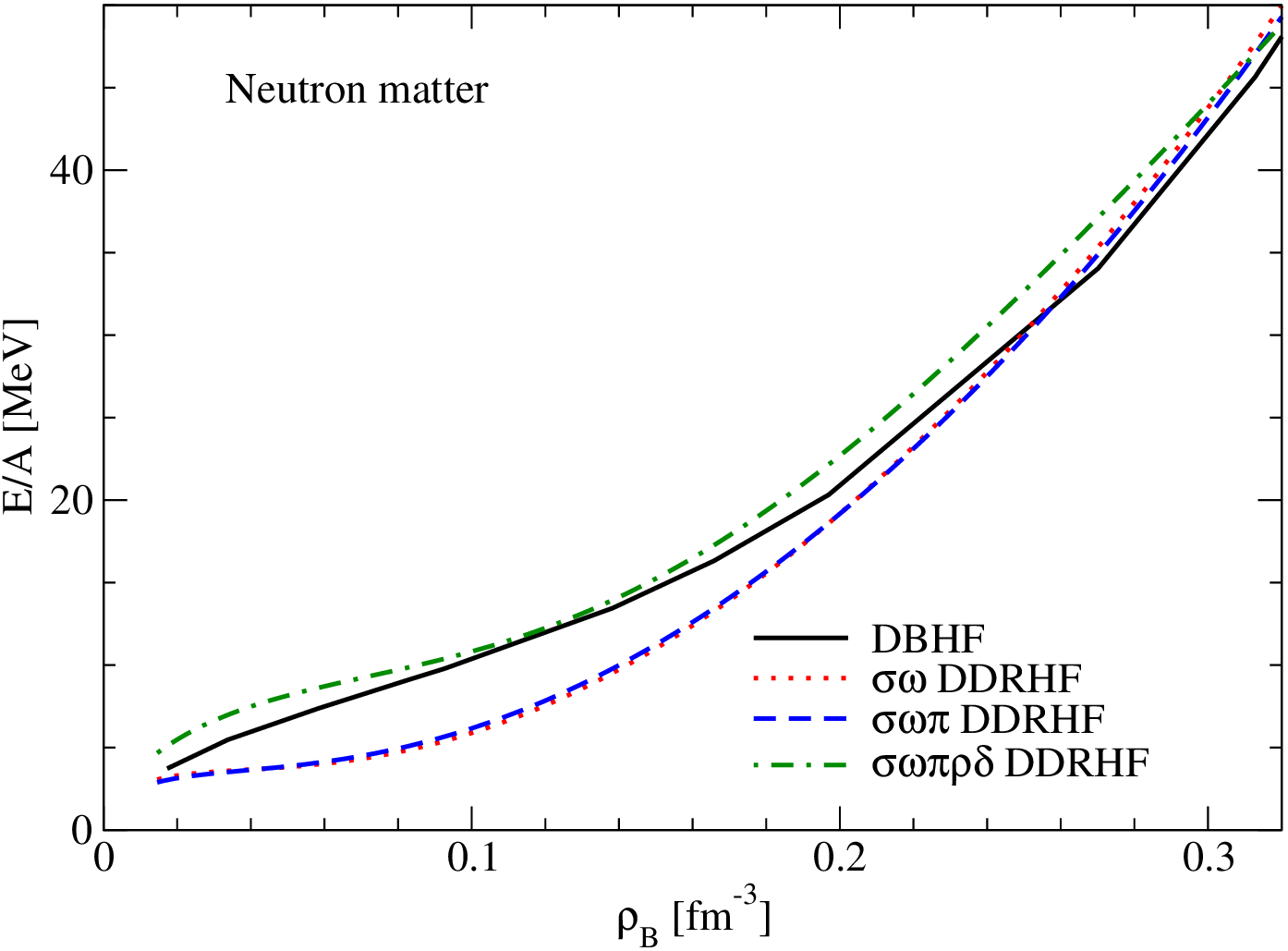}
\caption{(Color online) Energy per Nucleon is plotted as a function of the density for the $\sigma\omega$ (dotted), $\sigma\omega\pi$(dashed), and the $\sigma\omega\pi\rho\delta$(dashed-dotted) DDRHF model. In addition, the energy per nucleon of the DBHF approach in Ref.~\cite{vandalen:2007} (solid line) is given. Left: isospin symmetric nuclear matter. Right: pure neutron matter.}
\label{fig:binding}
\end{center}
\end{figure}

The self-energy components are needed to calculate the energy per nucleon as can be
seen from the Eqs.~(\ref{eq:EoverA})-(\ref{eq:potential}). The energy per nucleon
for isospin symmetric nuclear matter and pure neutron matter are displayed in
Fig.~\ref{fig:binding}. The various approximation schemes reproduce the DBHF results
for symmetric matter reasonably well at medium and larger densities. At densities
below 0.8 fm$^{-3}$ all parameterizations tend to underestimate the binding energy
per nucleon calculated in the DBHF approach. The $\sigma\omega$ model is less
attractive than the DBHF and the other approaches at larger densities. This is
partly due to the weak spatial vector component but also due to the fact that the
difference between the self-energy components  $\Sigma_s$ and $\Sigma_0$ tends to be
too small at low momenta in this approach as can be seen in
Fig.~\ref{fig:momselfenergy}. 
\begin{table}
\begin{center}
\begin{tabular}{|ll|cc|ccc|c|}
\hline
&&\multicolumn{2}{c|} {Hartree} & \multicolumn{3}{c|} {DDRHF} & DBHF\\
	&& $\sigma\omega$ & $\sigma\omega\pi\rho\delta$ & $\sigma\omega$  & 
	  $\sigma\omega\pi$  & $\sigma\omega\pi\rho\delta$   &  \\
\hline	  
$\rho_0$ &       [fm$^{-3}]$ & 0.1572 \ &  0.1613 & 0.1508 \ &  0.1623 \ & 0.1583   & 0.181 \\
$E/A$   &[MeV]	            & -14.44  \  & -14.76  & -13.56 \ &  -15.25 \ & -16.34  & -16.15 \\
\hline
\end{tabular}
\end{center}
\caption{\label{table:nm_prop}
Saturation properties of nuclear matter for the various models compared to the DBHF results
of Ref.~\cite{vandalen:2004b,vandalen:2007}.
The quantities listed include the saturation density $\rho_0$ and the
binding energy $E/A$ at saturation density.}
\end{table}

Therefore,  at saturation density nuclear matter is too weakly bound  as can be
seen in Table.~\ref{table:nm_prop}. In addition, the saturation density is shifted
to lower densities compared to the original DBHF results. This shift of the
saturation density to a lower density than in the DBHF approach is also observed in
the other DDRHF models.  

For pure neutron matter, one finds in Fig.~\ref{fig:binding} that the isovector
mesons $\rho$ and $\delta$ are important to describe the DBHF results at low
densities in particular. The influence of these mesons gets smaller at high
densities, which is already indicated by the density-dependence of the coupling
constants shown in Fig.~\ref{fig:isovector}.

\subsection{Finite Nuclei}
For the study of finite nuclei, we account for the density dependent correlation
effects in a relativistic HF calculation by employing the coupling constants
calculated at the local density. The density profile $\rho_B(r)$ is determined from
the result of the relativistic HF calculation in a self-consistent
manner~\cite{fritz:1994}. Furthermore, the rearrangement
self-energy contribution is taken into account, since it is important to get
appropriate single-particle energies and wave functions in finite nuclei. 
To solve the Dirac equation for finite nuclei in coordinate space, the radial 
functions are expanded in a discrete basis of spherical Bessel functions. This
discrete basis forms a complete orthonormal basis in  a sphere of radius D. This
radius D is chosen to be 30 fm, which is large enough to ensure that the results 
for the bound single-particle states are independent on D.   

\begin{table}
\begin{tabular}{|c|cc | cc| cc | cc | cc| cc|}
\hline
Nucleus    &   \multicolumn{2}{c|}{ Hartree } &  \multicolumn{2}{c|} {Hartree}
  &   \multicolumn{2}{c|} {DDRHF} &   
 \multicolumn{2} {c|}{DDRHF} &   \multicolumn{2}{c|}
 {DDRHF} &\multicolumn{2}{c|} {Exp.}\\
   &   \multicolumn{2}{c|} {$\sigma\omega$}  &  \multicolumn{2}{c|}
   {$\sigma\omega\pi\rho\delta$}
 &   \multicolumn{2}{c|} {$\sigma\omega$} &   
 \multicolumn{2} {c|}{$\sigma\omega\pi$} &   \multicolumn{2}{c|}
 {$\sigma\omega\pi\rho\delta$} & \multicolumn{2}{c|}{ } \\&  E/A \ \ & $r_{CD}$ \
 \ & E/A \ \ &  $r_{CD}$ \ \ & E/A  &$r_{CD}$& E/A & $r_{CD}$& E/A & 
$r_{CD}$& E/A & $r_{CD}$ \\
\hline
$^{16}O$   &   -6.49 &   2.75    & -6.65&   2.72 &  -6.00  &  2.77     &  -7.63   &  2.67     & -8.91   &2.62 & -7.98 & 2.74\\
$^{40}Ca$  &     -7.10 &  3.47    & -7.29&   3.43 & -6.54 &    3.49     &  -8.08 &     3.39     & -9.18&  3.34 & -8.55 &  3.48\\
$^{48}Ca$  &     -7.29 & 3.51    & -7.20&  3.47& -6.64 &     3.53     &  -7.67 &    3.43     & -8.81&   3.40 & -8.67 &  3.47\\
$^{90}Zr$  &    -7.20 &    4.29    & -7.26 &4.23  & -6.56  &  4.32     &  -7.70 &  4.20     & -8.83&   4.16 & -8.71  &  4.27\\
\hline
    \end{tabular}
\caption{Results for the binding energies per nucleon in MeV and the charge radii in fm for 
the various DDRH and DDRHF models. The calculated energies
have been corrected by subtracting the spurious energy of the center of mass
motion. The experimental values are taken from Ref.~\cite{hofmann:2001}.}
\label{table:Eb}
\end{table}

We have investigated some closed-shell nuclei with our DDRHF model. The 
results obtained for the binding energy and charge radius are presented in
Table~\ref{table:Eb}. The center of mass correction is included in the displayed
binding energies and the charge radii have been evaluated from the proton density,
assuming a radius of 0.8 fm for the charge radius of the proton. 

The $\sigma\omega$ DDRHF model yields with too little binding for all nuclei considered.
This outcome is expected from the nuclear matter results, since this model is too
repulsive in symmetric nuclear matter as can be seen in Fig.~\ref{fig:binding}. The
explicit inclusion of the $\pi$-exchange improves the results substantially. The  
$\sigma\omega\pi\rho\delta$ model yields too much binding energy for the  light
nuclei, whereas a good agreement to the empirical values is obtained for the heavier
nuclei. Furthermore, all models predict charge radii, which are in good agreement
with experimental values.

\begin{table}
\begin{tabular}{|c | l  c  |  l  c| l  c  | l c |}
\hline

    &  $\sigma\omega$ \ \  \ \ \ \ \ \ \ \  \   & &  $\sigma\omega\pi$  \ \  \ \ \ \
    \  &  & $\sigma\omega\pi\rho\delta$\ \ \  &  & exp. \ \  \ \ \ \ \  \ \ \ &\\
\ \ \ \ \ Orbital \ \ \ \ \ & Proton  & Neutron &  Proton & Neutron & Proton & Neutron & Proton & Neutron\\
\hline
 $^{16}O$   & & & &  & & & &\\
\hline
$0s_{1/2}$  & -27.62 & -31.86 & -34.64 & -39.05 & -37.06 & -41.50 & -44$\pm$7& -47\\
$0p_{3/2}$  & -13.16 & -17.09 & -16.55 & -20.65 & -18.73 & -22.90 & -18.451&   -21.839 \\
$0p_{1/2}$  & -9.35  & -13.20 & -12.75 & -16.78 & -14.23 & -18.34 & -12.127 & -15.663 \\
    \hline
$^{40}Ca$   & & & &  & & & & \\
\hline
$0s_{1/2}$  & -33.13 & -41.45 & -41.28 & -49.91 & -43.15 & -51.77 & -49.1$\pm$12 &\\
$0p_{3/2}$  & -22.11 & -30.06 & -27.79 & -36.00 & -30.19 & -38.43 & -33.3$\pm$6.5& \\
$0p_{1/2}$  & -19.55 & -27.45 & -25.30 & -33.47 & -27.42 & -35.61 & -32$\pm$4&\\
$0d_{5/2}$  & -10.71 & -18.29 & -13.62 & -21.45 & -15.72 & -23.64 & -14.2$\pm$2.5&  -21.30\\
$1s_{1/2}$ & -6.70  & -14.23 & -8.95  & -16.78 & -9.31  & -17.33 & -10.850& -18.10\\
$0d_{3/2}$  & -6.61  & -14.08 & -9.50  & -17.23 & -11.01 & -18.85 & -8.325&  -15.641    \\
    \hline
$^{48}Ca$   & & & &  & & & &\\
\hline
$0s_{1/2}$   & -37.17 & -42.35 & -44.52 & -50.90 & -48.38 & -52.64 & &\\
$0p_{3/2}$   & -27.25 & -31.13 & -32.05 & -36.84 & -36.75 & -39.03 & & \\
$0p_{1/2}$   & -25.18 & -29.51 & -29.80 & -36.46 & -34.10 & -38.50 & &\\
$0d_{5/2}$   & -16.32 & -19.48 & -18.38 & -22.09 & -23.22 & -24.10 & -20$\pm1$&  \\
$1s_{1/2}$  & -11.26 & -15.57 & -13.03 & -18.45 & -15.84 & -19.08 & -15.8& \\
$0d_{3/2}$   & -12.55 & -16.18 & -14.47 & -20.93 & -18.68 & -22.44 & -15.3&      \\
$0f_{7/2}$   & (-4.90)& -7.89  & (-4.55)& -7.82  & (-8.75) &-8.95 & &\\
    \hline
    \end{tabular}
\caption{Single particle energies for the orbital levels of $^{16}O$, $^{40}Ca$, and $^{48}Ca$ nucleus
derived from RHF calculations using various models.
The experimental values are taken from Ref.~\cite{fritz:1994,coraggio:2003}.}
\label{table:spe}
\end{table}
Single particle energies for the orbital levels of the $^{16}O$, $^{40}Ca$, and $^{48}$Ca nucleus
are presented in Table~\ref{table:spe} for the various models. The $\sigma\omega$ DDRHF model predicts absolute values of single-particle energy,
which are too small. The single particle energies of the  $\sigma\omega\pi$ and $\sigma\omega\pi\rho\delta$ model are in good agreement with experimental data. They are in better agreement with experimental values than the energy levels calculated in Refs.~\cite{fritz:1994,shi:1995}, which lie to deep.
However, the order of the 1s-shell and 0d-shell deviates from experiment. The spin-orbit splitting plays a key role in the ordering of these shells.    
\begin{table}
\begin{tabular}{|c|c   | c | c  | c  |}
\hline
Nucleus    &   \ \ \ \ \ \   $\sigma\omega$  \ \ \ \ \ \           & \ \ \ \ \    $\sigma\omega\pi$ \ \ \ \ \ &        \  \  \    $\sigma\omega\pi\rho\delta$ \ \ \ & \ \ \ \  \    exp. \ \  \ \ \    \\
\hline
$^{16}O$   &    3.81 &    3.8          & 4.5         & 6.3 \\
$^{40}Ca$  &    4.1 &    4.12          & 4.71        & 7.2  \\
$^{48}Ca$  &    3.77 &   3.91          & 4.54        & 4.3   \\
$^{90}Zr$  &    1.03  &   1.22          & 1.51        & 1.5  \\

\hline
    \end{tabular}
\caption{Spin-orbit splittings of protons for the 0p-shell in $^{16}O$, the 0d-shell in $^{40}Ca$ and $^{48}Ca$, and the 1p-shell in $^{90}Zr$. The experimental values are taken from Ref.~\cite{shi:1995}.}
\label{table:spinorbit}
\end{table}
The spin-orbit splittings of some nuclei are given in Table~\ref{table:spinorbit}. It can be seen  that in all the presented DDRHF models the spin-orbit splittings in $^{16}O$ and $^{40}Ca$ are smaller than the experimental values. However, the agreement is better in $^{48}Ca$ and in $^{90}Zr$, in particular
for the  $\sigma\omega\pi\rho\delta$ model. 
In works of Ref.~\cite{fritz:1994,shi:1995}, it was reported that the $\pi$ exchange reduced the spin-orbit splitting. We find as can be seen in Table~\ref{table:spinorbit} that the inclusion of the $\pi$  has no noticeable effect on the spin-orbit splitting in light symmetric nuclei, whereas in heavier asymmetric nuclei  a small increase in the spin-orbit splitting can be observed. However, the inclusion of the isovector $\rho$ and $\delta$ meson leads to a clear increase of the spin-orbit splitting in as well light as heavier nuclei.

\begin{figure}[!ht]
\begin{center}
\includegraphics[width=0.6\textwidth] {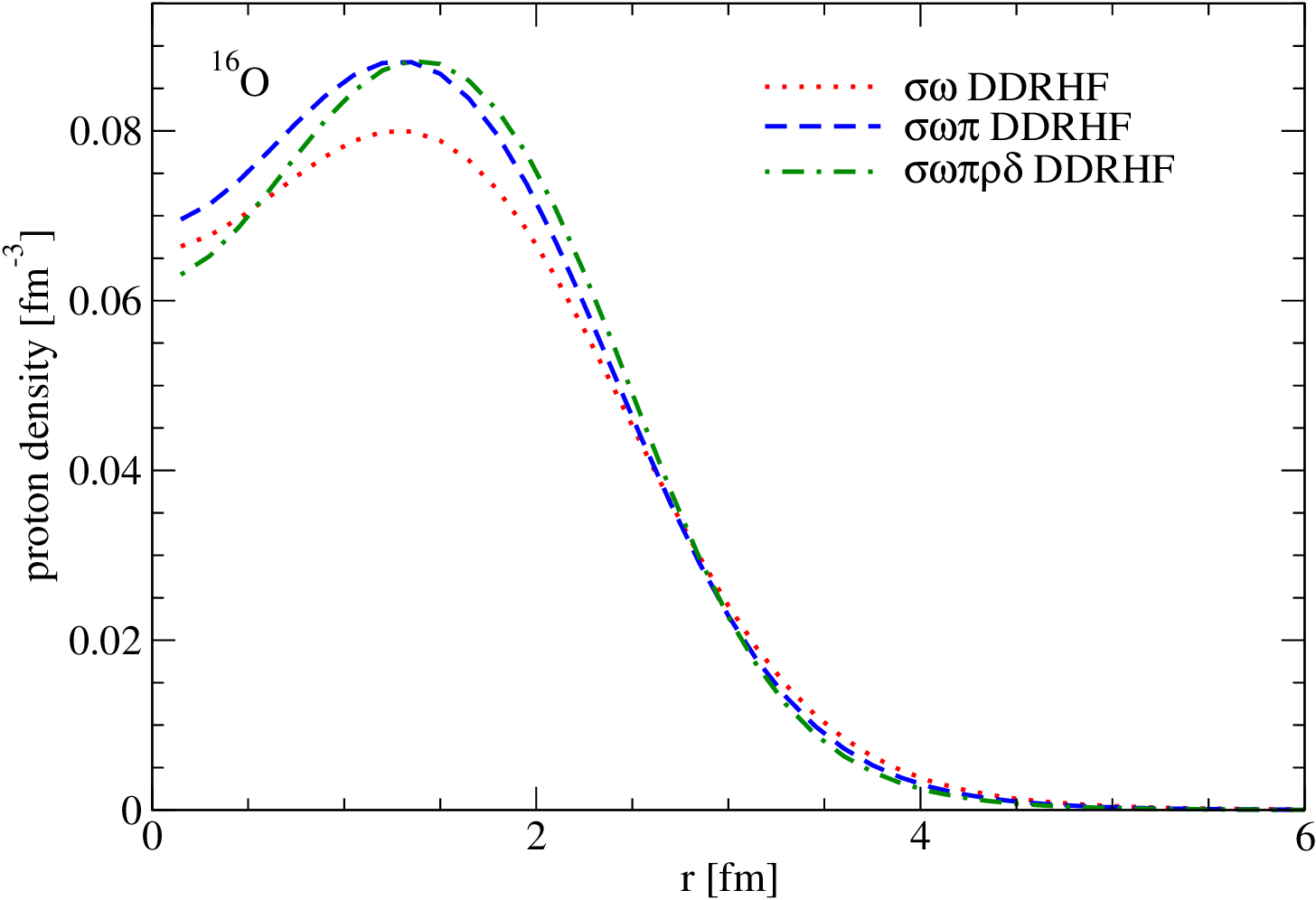} 
\caption{(Color online) Proton density distribution of $^{16}O$ nucleus for the $\sigma\omega$ (dotted), $\sigma\omega\pi$(dashed), and the $\sigma\omega\pi\rho\delta$(dashed-dotted) DDRHF model.}
\label{fig:pdensO16}
\end{center}
\end{figure}
Fig.~\ref{fig:pdensO16} shows the proton density distribution of the $^{16}O$ nucleus. It is found that in the $\sigma\omega$ model the proton density in the interior
is smaller than in the other two models and has a longer tail. Therefore, it has a larger charge radius as can be seen in Table~\ref{table:Eb}. Furthermore, the smaller maximum density in the $\sigma\omega$ model compared to the other models can be expected from its smaller saturation density in nuclear matter.

\section{Summary and Conclusion}
\label{sec:S&C}

Although Dirac-Brueckner-Hartree-Fock (DBHF) calculations
have been quite successful in describing nuclear matter, corresponding DBHF
calculations have not yet been performed for finite nuclei as such calculations
still seem to be too complex. Various attempts have already been made to
approximate such DBHF calculations by employing some kind of local density
approximation by parameterizing the results of DBHF calculations of nuclear matter
in terms of either a Dirac Hartree (DDRH) model or a Dirac Hartree Fock (DDRHF)
model with density-dependent coupling constants for various mesons considered.

It has been the aim of this investigation to compare these different models. 
The parameters of these DDRH and DDRHF models were all fixed to reproduce at each
density the same scalar $\Sigma_s$ and time-like vector component $\Sigma_0$
component of the self-energy 
obtained in the DBHF calculation for nucleons with momentum equal to the Fermi
momentum $k_F$.  

While the DDRH approach yields no momentum dependence of the
self-energies, the DDRHF model reproduces the qualitative features of the momentum
dependence obtained in the DBHF calculations. However, only 50 percent of this
momentum dependence can be related to the Fock exchange terms. The remaining part of
the momentum dependence in DBHF is due to the non-locality of the DBHF $G$-matrix.
It remains as a challenge for further investigations to account for this
non-locality in such a way that the complete energy- and momentum-dependence of the 
self-energy is described in nuclear matter using a representation, which can be
transferred to calculations of finite nuclei. 

The spatial vector component of the nucleon self-energy $\Sigma_v$  of the DBHF is
dominated by the $\pi$-exchange term and therefore can be accounted for in DDRHF
models, which include the $\pi$ explicitly. The exchange of isovector mesons $\rho$
and $\delta$ is important to describe the differences of correlation effects in
isospin $T=0$ and $T=1$ channels. Therefore a DDRHF model accounting for
$\sigma,\omega,\pi, \rho$ and $\delta$ exchange is required to reproduce the
features of DBHF calculations for symmetric and asymmetric nuclear matter.

Fair agreement with empirical data is obtained when this parameterization of the
full $\sigma\omega\pi\rho\delta$ model is employed in Dirac Hartree Fock
calculations of finite nuclei with density dependent coupling constants for light
and heavy nuclei. The parameterization of the coupling constants is presented in
form which makes it accessible for other users.

\begin{acknowledgments}
This work has been supported by
the Deutsche Forschungsgemeinschaft (DFG) under contract no. Mu 705/5-2.
\end{acknowledgments}



\end{document}